\documentclass[reprint,aps,pre,superscriptaddress]{revtex4-1}

\usepackage[utf8]{inputenc}
\usepackage[T1]{fontenc}

\usepackage{natbib}
\usepackage{graphicx}
\usepackage{dcolumn}
\usepackage{bm}
\usepackage{xcolor}
\usepackage{amsmath}
\usepackage{comment} 
\usepackage{amssymb}
\usepackage{mathrsfs} 
\usepackage[normalem]{ulem}

\renewcommand{\L}{\mathscr{L}} %
\renewcommand{\d}{\mathrm{d}} 
\renewcommand{\Re}{\operatorname{Re}} 
\renewcommand{\Im}{\operatorname{Im}} 

\newcommand{\xd}{x_{\downarrow}}
\newcommand{\xu}{x_{\uparrow}}

\date{March 2019}

\begin{document}
\title{Mean Field Control for Efficient Mixing of Energy Loads}

\author{David M\'etivier}
\affiliation{Theory Division \& CNLS, LANL, Los Alamos, NM 87544, USA}
\author{Michael Chertkov}
\affiliation{Program in Applied Mathematics, Department of Mathematics, University of Arizona,
        Tucson, AZ 85721, USA \& Theory Division, LANL, Los Alamos, NM 87544, USA}

\begin{abstract}
We pose an engineering challenge of controlling an \emph{Ensemble of Energy Devices} via coordinated, implementation-light and randomized on/off switching as a problem in \emph{Non-Equilibrium Statistical Mechanics}.  We show that \emph{Mean Field Control} with nonlinear feedback on the cumulative consumption, assumed available to the aggregator via direct physical measurements of the energy flow, allows the ensemble to recover from its use in the \emph{Demand Response} regime, i.e. transition to a statistical steady state, significantly faster than in the case of the fixed feedback. Moreover when the nonlinearity is sufficiently strong, one observes the phenomenon of ``super-relaxation'' -- where the total instantaneous energy consumption of the ensemble transitions to the steady state much faster than the underlying probability distribution of the devices over their state space, while also leaving almost no devices outside of the comfort zone.  
\end{abstract}

\maketitle

\section{Introduction}
This manuscript is devoted to analysis and acceleration of mixing within a large ensemble of energy-consuming loads.
The ensemble is assumed controlled by a middle-ground, profit-seeking entity coined ``aggregator.'' The aggregator is set to balance between the following two, generally conflicting, sides: (a) making the ensemble responsive and available for the system operator's demand response (DR) \cite{DR} action, which is aimed to balance, with the least delay possible, all generation and loads originating from resources and customers which cannot be controlled efficiently, and (b) avoiding putting the controlled ensemble-forming customers/loads at risk of spending too much time outside of their comfort zone. The ensemble control can also be viewed as aiding system operator by introducing a virtual battery \cite{virtual_battery,hao2015aggregate} to balance the system in real time, see \cite{desrochers2019real} for real word implementation or \cite{meyn} for a popularized presentation. 

We consider the following basic setting discussed extensively (with some variations) in multiple studies \cite{79MB,79CD,81IS,84CM,2013ETH,15PKL,16BMa,16BMb,2018CCD}: 
    1) Ensemble is formed by aggregator from similar loads. 
    2) An air conditioner load in an apartment is considered. When a load does (does not) consume electricity, i.e. when it is switched on (off), temperature in the apartment decreases (increases).
    3) By default, i.e., in their normal operational mode, when no signal from the aggregator arrives, loads function according to the so-called ``bang-bang'' control --- air conditioner is switched on (off), temperature reaches $x_\downarrow$ ($x_\uparrow$), and then the air conditioner is switched off  (on). Here we assume that outside temperature and capacity of the air conditioner are such that the load never reaches a point of dynamic equilibrium thus it is always cycling in its mixed (temperature + on/off) phase space. 
    4) Aggregator controls ensemble by communicating the same control signal to all loads, and the signal overrides default controls.

Already, in the earliest studies on the subject a significant complication, coined ``cold load pick up'' \cite{79MB,79CD,81IS,11CH}, was revealed --- a period of extensive use of the ensemble in the DR leads to synchronization of load dynamics along the cycle, turning the ensemble from a useful, flexible reserve to an entity whose energy consumption oscillations should be compensated by some other means. Therefore, every period of use, when ensemble is utilized for DR, should follow a grace period, during which ensemble loads are not used for DR, to ensure graceful transition to a well-mixed steady state. Natural stochasticity, which would eventually mix the loads, is typically weak. Then a plausible strategy for the aggregator becomes to relax the default (bang-bang) control policy to achieve a faster mixing while also keeping number of loads outside of the comfort zone to a minimum. As shown in \cite{12AK,2013ETH,16BMb,17CC,18MLC} randomization of the control policy \cite{12AK,16BMb,17CC} and diversification of loads by types within the ensemble \cite{2013ETH,18MLC} both help to accelerate mixing. In this manuscript we continue to focus on the task of acceleration however exploring a new, so-called \underline{Mean-Field Control} (MFC) \cite{huang2006,Lasry2007,2011GLL,13NC}, option. Our version of the MFC is built on the assumption that the aggregator measures and communicates to the loads  the total energy consumption of the ensemble (broadcasting the same signal to all and without delays), and then the loads use a {\bf NONLINEAR} randomization policy depending on the instantaneous averaged over the ensemble consumption.  Significant for massive adaptation of the DR technology,  this scheme is {\bf SECURE} and {\bf PRIVATE} because the aggregator does not require access to the consumption of individual loads.  The aggregator only needs access to the meter located at the head of the line feeding all the loads --- this is under assumptions that (a) loads are geographically collocated, (b) loads are connected to the same energy source (feeder),  (c) contribution of non-ensemble loads is known and excluded from the energy balance, (d) energy losses within the feeder are measured/known.
 
\subsection{Model}
 
\begin{figure}
    \includegraphics[width=0.485\textwidth]{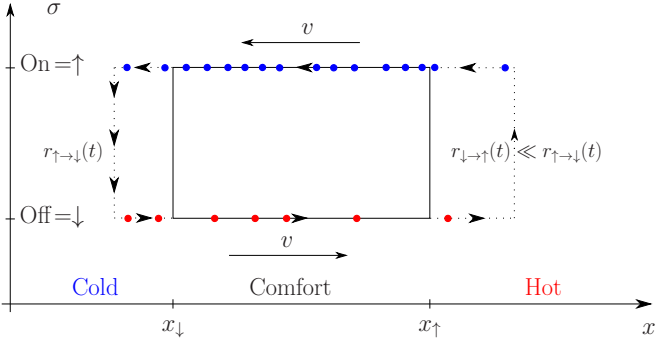}
    \caption{ \label{fig:x-sigma} Schematic illustration of an instantaneous device distribution and dynamics in the $(x,\sigma)$ phase space, where $x$ is the (continuous) temperature and the discrete (two-level) designation $\sigma=\uparrow,\downarrow $  marks whether the device (air conditioner) is switched on, $\sigma=\uparrow$, or off, $\sigma=\downarrow$. Instantaneous placement of cold (blue) and hot (red) loads at some $t$ illustrates an inhomogeneous case where a majority of the devices are on. In this case MFC makes the transition rate from off-to-on, $r_{\downarrow\to \uparrow}$, higher than the on-to-off rate, $r_{\uparrow\to \downarrow}$, thus boosting equalization of the two densities (in comparison with the case where the two rates are equal).}
\end{figure}

To turn this setting into stochastic, dynamic, nonlinear model we assume that all devices follow the same rules described in terms of the device's state, $(x(t),\sigma(t))$,  where the continuously valued $x(t)$ is  the temperature measured inside the apartment associated with the device 
and the binary $\sigma(t)$ shows if the device  is in the switch-on, $\sigma(t)=\uparrow$, or switch-off, $\sigma(t)=\downarrow$, mode respectively at time, $t$. The state of the device evolves in time, according to the following rules:
\begin{eqnarray}
 && \frac{dx}{dt}=\left\{
    \begin{array}{cc} 
        -v, & \sigma=\uparrow
        \\ v, & \sigma=\downarrow
    \end{array}\right. , \quad \forall t\label{eq:basic_x_model}\\ 
&& \sigma(t+dt)\!=\!\left\{\!\begin{array}{cc} \sigma(t), & x(t)\in [x_\downarrow;x_\uparrow]
    \\ -\sigma(t), & x(t)<\xd \mbox{ with prob. } r_{\uparrow\to \downarrow} \d t 
	\\ -\sigma(t), & x(t)>\xu\mbox{ with prob. }  r_{\downarrow \to \uparrow}\d t     
    \end{array}\right.  \label{eq:sigma_model}
\end{eqnarray}
where in Eq.~(\ref{eq:basic_x_model}), $\pm v$ describe the instantaneous rate of the temperature change, depending on if the air conditioner is switched on or off and $[x_\downarrow;x_\uparrow]$ is the preset temperature interval where devices do not switch, while $r_{\uparrow\to \downarrow}$ and $r_{\downarrow\to \uparrow}$ in Eq.~(\ref{eq:sigma_model}) are the two distinct rates of switching of the device's mode from on-to-off and off-to-on, respectively, communicated by the aggregator.  Fig.~(\ref{fig:x-sigma}) illustrates an instantaneous distribution (snapshot) of multiple devices (dots) distribution in the $(x,\sigma)$ space, and their dynamics (arrows) according to Eqs.~(\ref{eq:basic_x_model},\ref{eq:sigma_model}).

In the thermodynamic limit, when the number of devices, $N$, becomes infinite, evolution of the distribution vector monitoring fraction of the devices staying in the on and off states at temperature $x$ and time $t$, $P(x|t)=(P_{\uparrow}(x|t),P_{\downarrow}(x|t))^\intercal$, is governed by the system of coupled Nonlinear Fokker--Planck (NFP) equations following directly from the model definition Eqs.~(\ref{eq:basic_x_model}, \ref{eq:sigma_model}). (See also \cite{18MLC} and references therein) 
\begin{eqnarray}
    && \partial_t P(x|t) =\mathcal{L} P(x|t),\label{eq:FP}\\
    && \mathcal{L} P \doteq 
        \begin{pmatrix} 
        v & 0
        \\ 0 & -v
        \end{pmatrix}\partial_x
       P  -\begin{pmatrix} r_{\uparrow\to\downarrow} & - r_{\downarrow\to\uparrow}\\ -r_{\uparrow\to\downarrow} & r_{\downarrow\to\uparrow}\end{pmatrix}P. \label{eq:FP_full}
\end{eqnarray}
Notice that a more general modeling should account for device-specific stochastic effects in Eq.~(\ref{eq:basic_x_model}) resulting in additional diffusion contributions to the NFP Eqs.~(\ref{eq:FP_full}). We choose not to include these terms because their effect is expected to be much smaller than one of the nonlinear terms we are focusing our attention in the following. (This latter statement came from extensive numerical analysis accounting for the stochastic effects, which we choose not to include in the manuscript to avoid unnecessary complications.)
Properly normalized solution of Eq.~(\ref{eq:FP}) satify:
\begin{eqnarray}
N_\uparrow(t)+N_\downarrow(t)=1,\quad N_{\uparrow,\downarrow}(t)\doteq \int  P_{\uparrow,\downarrow}(x|t) \,\d x,
        \label{eq:densities}
\end{eqnarray}
where $N_{\uparrow,\downarrow}(t)$ counts proportions of devices that are switched on and off, respectively. 

In \cite{18MLC} the rates $r_{\uparrow\to \downarrow}$ and $r_{\downarrow\to \uparrow}$  were considered equal to each other and moreover constant.  The main point of this manuscript is to explore the possibility of controlling the two rates not only independently but also, and most importantly, in a nonlinear way. We consider here the nonlinearity of a special kind --- associated with the \textbf{MFC} (often refered to in the literature as the mean-field game \cite{huang2006,Lasry2007,2011GLL,13NC}),  where the rates are dependent on the instantaneous state of the entire ensemble.  Specifically,  we consider the possibility for the rates to depend on the instantaneous value of the total number of the devices which are switched on. This specific choice is dictated by our ability to measure total number of devices directly through physical observation of the total energy consumption of the ensemble. This suggests to 
consider nonlinear feedback on the instantaneous aggregated consumption, $N_{\uparrow}$:
\begin{eqnarray}
r_{\uparrow\to \downarrow}(N_\uparrow(t),x) & = & 
\begin{cases}
g\left (N_\uparrow(t)\right),\quad x<\xd\\
0 \qquad\qquad\text{otherwise}
\end{cases},
\label{eq:rup}\\
r_{\downarrow\to \uparrow}(N_\uparrow(t),x) & = &
\begin{cases}
g\left(1-N_\uparrow(t) \right),\quad x>\xu\\
0  \qquad\qquad\text{otherwise}
\end{cases},
\label{eq:rdown}
\end{eqnarray}
where the functional form of $g(N)$ can be tuned (by the system operator) to achieve the desirable effect of efficient mixing. In the following we will limit our analysis to the following specific ``sufficiently nonlinear" form of 
\begin{equation}
g(N) = r (N/N^{(\rm{st})})^s, \label{eq:s}
\end{equation}
where $s$ measures the degree of nonlinearity, and $r$ is the rate at the equilibrium, when $N=N^{(\rm{st})}$. For $s = 0$, $g(N) = r$ is a constant and the system becomes linear. It is important to stress that Eq.~(\ref{eq:FP}) should also be supplemented by properly defined boundary conditions, e.g. accounting for discontinuity of the rates dependence on $x$. These conditions are discussed in  Section \ref{subsec:linearization}).

To clarify intuition behind setting up non-constant and nonlinear rates (\ref{eq:rup},\ref{eq:rdown}), let us assume that  due to a perturbation there appears to be more devices in the  ``on'' state, $N_\uparrow(t)>N^{(\rm{st})}$. Then the switching off rate dominates,
$r_{\uparrow\to\downarrow} > r_{\downarrow\to\uparrow}$, leading to decrease of $N_\uparrow$,  therefore suggesting that the feedback mechanism will accelerate the desired relaxation, $N_\uparrow(t)\to N^{(\rm{st})}$.

Notice similarity of the nonlinearity in the rate with the mean-field coupling analyzed in plasma physics \cite{landau_vibrations_1946,vlasov_vibrational_1968,mouhot_landau_2011} and synchronizing networks \cite{kuramoto_self-entrainment_1975,kuramoto_cooperative_1984,strogatz_kuramoto_2000},  where the nonlinearity is associated with the names of Vlasov and Kuramoto, respectively. 


\section{Analysis of the Nonlinear Fokker--Planck Equation}
\label{sec:NFP}

\subsection{Steady State.} 
Analyzing time-independent solution of the NFP system of Eqs.~(\ref{eq:FP},\ref{eq:FP_full},\ref{eq:rup},\ref{eq:rdown}) one can explicitly compute the steady state, 
$P^{\rm{(st)}}(x)=(P_{\uparrow}^{\rm{(st)}}(x),P_{\downarrow}^{\rm{(st)}}(x))$,
and with Eq.~\eqref{eq:densities} finds that $N_\uparrow^{(\rm{st})}=N_\downarrow^{(\rm{st})}=1/2$. Then the switching rates are also equal to $r\doteq g(1/2)$ and fraction of devices $N_{\rm{out}}(t)$ residing outside the comfort zone 
\begin{eqnarray}
N_{\rm{out}}(t)\doteq \int_{x<\xd \& x>\xu}  P_{\uparrow}(x|t) \,\d x,
\end{eqnarray}
is $N_{\rm{out}}^{\rm{(st)}}=(1+r \tau/4)^{-1}$, where $\tau\doteq2(\xu-\xd)/v$ is the cycling period in the $r\to\infty$ limit.


\subsection{Dynamics  around the Stationary Solution}
\label{subsec:linearization}

We are interested to analyze dynamics which results in relaxation to the stationary solution described in the preceding subsection. At the latest stages of this dynamical process, correction to the stationary solution for the probability density is small and thus governed by the following linearized version of Eq.~\eqref{eq:FP_full}  
\begin{flalign}
 P(x|t)&= P^{\rm{(st)}}(x)+\rho(x,t),~ \|\rho\|\ll \|P^{\rm{(st)}}\|,\label{eq:linearization}
 \\\rho(x|t)&\doteq  \begin{pmatrix} \rho_{\uparrow}(x|t)\\ \rho_{\downarrow}(x|t)\end{pmatrix}, 
\label{eq:chi++s}\\
\dfrac{\partial \rho}{\partial t} -&\L \rho =   rs\begin{pmatrix} -\chi_\uparrow \theta(x<x_\downarrow) & \chi_\downarrow \theta(x>x_\uparrow)\\ \chi_\uparrow \theta(x<x_\downarrow) & -\chi_\downarrow \theta(x>x_\uparrow) \end{pmatrix}P^{\rm{(st)}}(x) 
,\label{eq:lin_NFP}
\\ \L &\doteq \begin{pmatrix} v & 0\\ 0   & -v\end{pmatrix}\partial_x +
r\begin{pmatrix}-\theta(x<x_\downarrow) & \theta(x>x_\uparrow)\\ \theta(x<x_\downarrow) & -\theta(x>x_\uparrow)\end{pmatrix}, \label{eq:L-lin}
\\\chi(t) &\doteq \begin{pmatrix} \chi_{\uparrow}(t)\\ \chi_{\downarrow}(t) =-\chi_{\uparrow}(t)\end{pmatrix}\doteq \int \rho(x',t)\,\d x',
\label{eq:chi+s}
\end{flalign}
where 
$\theta(a)$ is unity if $a$ is true, and it is zero otherwise. 
Notice that when the right-hand side of Eq.~(\ref{eq:lin_NFP}) is replaced by zero, i.e. when $s$ introduced according to Eq.~(\ref{eq:s}) is set to zero, the resulting equation becomes the basic (linear) Fokker--Planck equation for which solutions were analyzed in previous work \cite{18MLC}. Looking for the solution of Eq.~(\ref{eq:lin_NFP}) in the form of a spectral series,
\begin{eqnarray}
\rho(x|t)=\sum_\lambda a_{\lambda} \exp(-\lambda t)\rho_\lambda(x),
\label{eq:spectr_exp}
\end{eqnarray}
where $a_\lambda$ are coefficients dependant on the initial condition, making substitution
$\chi(t)=\sum_\lambda a_\lambda\exp(-\lambda t)\chi_\lambda$ and considering $\chi_\lambda$ as a vector of fixed constants (which should be found later on from a consistency condition), one arrives at a linear inhomogeneous differential equation. Seeking for solution of the equation in the form of a sum of an arbitrary zero mode of  the operator, $\lambda+\L$, and a particular inhomogeneous solution, one derives 
    \begin{eqnarray} &&
    \rho_\lambda(x)=
              \begin{cases}
                  c_L e^{\frac{x (r-\lambda )}{v}}
                	\begin{pmatrix}
                		   1\\ 
                		   \frac{r}{r-2\lambda}
            		\end{pmatrix}
            	     , ~ x<\xd\\\\
            		\begin{pmatrix}
                		 c_{c_\uparrow}  e^{-\frac{\lambda  x}{v}}\\ 
                		 c_{c_\downarrow}  e^{\frac{\lambda  x}{v}}
            		\end{pmatrix}
            		,~\xd<x<\xu \\\\
            		c_R e^{\frac{-(r-\lambda)  x}{v}}
                	\begin{pmatrix}
                		   \frac{r}{r -2\lambda}\\ 
                		   1
            		\end{pmatrix}
            		,~ x>\xu
            	\end{cases}
    +\nonumber\\ &&
    rs\dfrac{\chi_{\lambda;\uparrow}}{N_Q} 
                \begin{cases}
                      \frac{e^{\frac{r (x-\xd)}{v}}}{\lambda}
                	\begin{pmatrix}
                		   1\\ 
                		   \frac{r+\lambda }{r-\lambda}
            		\end{pmatrix}
            	     , ~ x<\xd\\\\
            		\begin{pmatrix}
                		 0\\ 
                		 0
            		\end{pmatrix}
            		,~\xd<x<\xu \\\\
            		-\frac{e^{-\frac{r (x-\xu)}{v}}}{\lambda }
                	\begin{pmatrix}
                		   \frac{r+\lambda }{ r-\lambda }\\ 
                		   1
            		\end{pmatrix}
            		,~ x>\xu
            	\end{cases}.
    \label{eq:rho_lambda_sol}
    \end{eqnarray}
Requiring that the eigenvector Eq.~\eqref{eq:rho_lambda_sol} are continuous at the boundaries $x=x_{\downarrow,\uparrow}$ and then imposing the self-consistency condition $\chi_{\lambda;\uparrow}=\int \rho_{\lambda;\uparrow}(x)\,\d x$ one arrives at the following two families of equations that must be satisfied by $\lambda$:
\begin{flalign}
 &\frac{r e^{\frac{\lambda  \tau }{2}}}{r-2 \lambda } \left(\frac{4s r^2}{(r-\lambda ) \left(2 s r+(r-\lambda ) (r \tau +4)\right)}-1\right)=1, \label{eq:spectrum-}\\
 &\frac{r e^{\frac{\lambda  \tau }{2}}}{r-2 \lambda } =1, \label{eq:spectrum+}
\end{flalign}
where Eq.~(\ref{eq:spectrum+}) followed from the degenerate version of the continuity condition for the eigenvectors with $\chi_{\lambda;\uparrow}=0$. (Let us remind that $\tau\doteq2(\xu-\xd)/v$ is the cycling period in the case of the instantaneous bang-bang switching, $r\to\infty$, limit.)

\begin{figure}
            \includegraphics[width=0.485\textwidth]{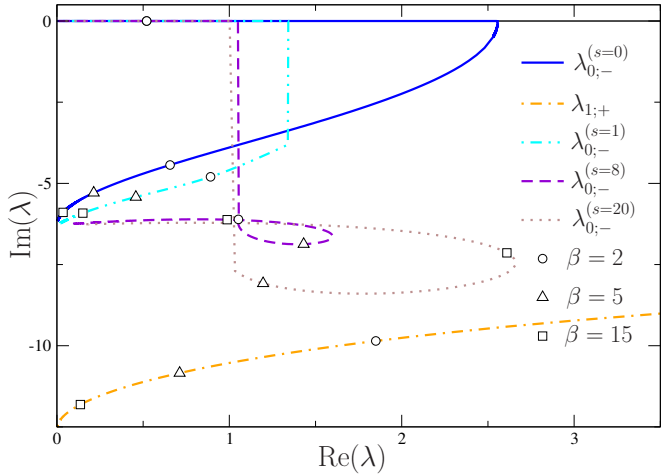}
             \caption{Multiple solutions of the spectral Eq.~(\ref{eq:spectrum-}) are visualized as curves in the 
             $\Re(\lambda),\Im(\lambda)$ plane as they change with $r$ when $\tau=1$. (Only lower half-plane is shown as the upper half-plane is symmetric.) Different curves represent the dominant eigenvalue of Eq.~\eqref{eq:spectrum+}, $\lambda_{1;+}$, at different values of $s$. The leading eigenvalue of Eq.~\eqref{eq:spectrum+}, $\lambda_{1;+}$, is also shown.  Dashed vertical lines correspond to the points where transition, from the regime where the relaxation is of the strict decay type (eigenvalue is real) to the regime where it is of the decay + oscillations type (eigenvalue is complex), occurs.
             }
    \label{fig:eigen}
\end{figure}

\begin{figure}
            \includegraphics[width=0.485\textwidth]{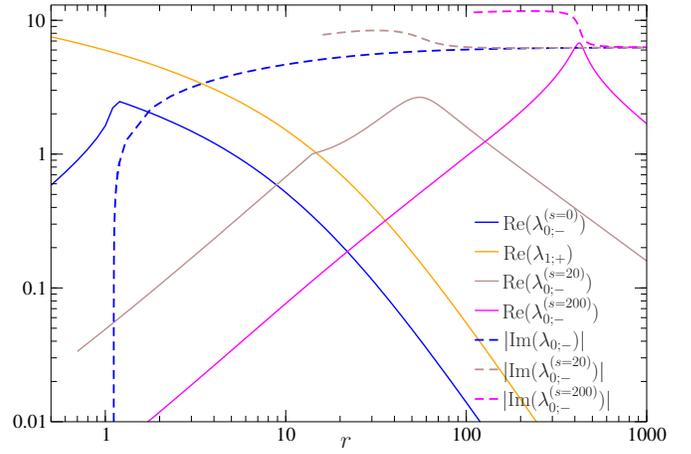}
             \caption{Real (solid line) and imaginary (dashed line) parts for solutions of the spectrum Eqs.~(\ref{eq:spectrum-}, \ref{eq:spectrum+}) evaluated for three values of the mean-field strength ($s=0,20,200$) and shown as a function of $r$ at $\tau=1$. 
             When, with $r$ increase, the real part of the mean-field eigenvalue crosses the $\Re(\lambda_{0;-}^{(s=0)})$ curve (marked blue), it signals a transition to the regime where the MFC results in a faster relaxation to the stationary value.  When $r$ is increased further, the mean-field eigenvalue can cross the $\Re(\lambda_{1;+})$ line, which is an indicator that $N_{\uparrow}(t)$ relaxes faster than any other expectations over $P(x|t)$ (i.e. super-relaxation). The end of the dashed curves (at sufficiently small $r$ and $s>0$) corresponds to the discontinuous transition from the relaxation + oscillation regime (at larger $r$) to the pure relaxation regime (at smaller $r$). The transition is continuous at $s=0$. We show the transition to oscillatory relaxation (when $\Im(\lambda)\neq 0$) for the different cases, which is continuous for $s=0$ and discontinuous otherwise.
             } 
    \label{fig:eigen_r}
\end{figure}

In the following we denote the discrete spectrum, emerging from solving Eq.~(\ref{eq:spectrum-}) and Eq.~(\ref{eq:spectrum+}) as ``$-$''-branch $k\in\mathbb{Z}:\quad  \lambda_{k;-}$ and ``$+$''-branch $k\in\mathbb{Z}^*:\quad  \lambda_{k;+}$, respectively.  A number of remarks are in order. First, let us stress that the two branches, represented by solutions of Eq.~(\ref{eq:spectrum-}) and Eq.~(\ref{eq:spectrum+}), coexist. Second, notice that the ``$+$''-branch, described by $\lambda_{k;+}$, (a) is mean-field nonlinearity--independent, and (b) does not contribute to $N_\uparrow$, which is our main object of interest, because as noted in \cite{18MLC}, $\int \rho_{\lambda_{k:+}}(x)\,\d x=0$. Third, the dependence of both branches on $x_{\uparrow,\downarrow}$ and $v$ is implicit via $\tau$. Fourth, all solutions of 
Eqs.~(\ref{eq:spectrum-}, \ref{eq:spectrum+}) result in non-negative $\lambda_{k;\pm}$, where $\lambda_{0,+}=0$, because it corresponds to the stationary state. Finally, fifth, at $s=0$ one recovers in Eq.~(\ref{eq:spectrum-}) the negative solution branch of the bare spectral problem (without the MFC) discussed in \cite{18MLC}. In this case, solutions of Eqs.~(\ref{eq:spectrum-},\ref{eq:spectrum+}) are given by $\lambda_{k;\pm}^{(s=0)} = 2\tau^{-1}\left (\beta- W_k(\pm \beta e^\beta)\right )$ for $k\in\mathbb{Z}$ where $\beta=r\tau/4$ and $W_k(x)$ is the Lambert-W function. In the following we will denote  dependency on $s$, for the ``$-$''-branch as $\lambda_{k;-}^{(s)}$, and omit it in the ``$+$''-branch which is not modified by the mean field. Moreover, as for the $s=0$ case, we will denote the leading eigenvalue (having the smallest real part) of the ``$-$''-branch by $k=0$ (and $k=-1$, its complex conjugate, when it becomes complex $\Im(\lambda_{0;-}^{(s=0)})\neq 0$). Similarly $\lambda_{\pm 1;+}$, which are always complex, will be the two leading eigenvalues of the ``$+$''-branch. $|k|>1$ will index eigenvalues with larger real part. 
Solutions of Eq.~(\ref{eq:spectrum-}), correspondent to the dominant eigenvalue, are shown in Fig.~\ref{fig:eigen} as curves in the  $\Re(\lambda),\Im(\lambda)$ as they change with $r$ at $\tau=1$.
The Fig.~\ref{fig:eigen_r}, is another representation of the spectrum on which one can see the various regime of the dynamics i.e. a) MFC is slower than no MFC: $\Re(\lambda_{0;-}^{(s)})<\Re(\lambda_{0;-}^{(s=0)})$, b) MFC is faster than no MFC: $\Re(\lambda_{0;-}^{(s)})>\Re(\lambda_{0;-}^{(s=0)})$, c) Super-relaxation: $\Re(\lambda_{0;-}^{(s)})>\Re(\lambda_{1;+})$.
Contrary to the case with $s=0$, when $s>0$ the transition from a real to complex eigenvalue is discontinuous in Eq.~(\ref{eq:spectrum-}), as it can be seen with the dashed lines in Fig.~\ref{fig:eigen_r}.

\subsubsection*{Asymptotic Analysis of the ``$-$''-branch}
\label{subsubsec:asymptotic}

One finds through perturbative analysis that in the regime of $r\tau\gg 1$ the  eigenvalue of the negative branch with the lowest real part becomes
\begin{eqnarray} 
\lambda_{0;-}^{(s)}=\lambda_{0;-}^{(s=0)}  +\frac{8 s}{r\tau^2}+O\left(\frac{1}{\tau^3 r^2}\right).
\end{eqnarray}
Hence, at $s>0$ and large $r\tau$, the correction (to the bare solution) is always real and positive, thus leading to a monotonic increase of the relaxation rate, $\Re(\lambda_{k;-}^{(s)})$, with an increase in the (level of) nonlinearity while also keeping the oscillatory contribution, $\Im (\lambda_{k;-}^{(s)})$, intact (in the leading order). The effect is clearly seen in Fig.~\ref{fig:eigen_r} showing numerical solution of Eqs.~(\ref{eq:spectrum-}, \ref{eq:spectrum+}). 

Another interesting regime is the regime of a weak nonlinearity corresponding to $s\ll 1$ and fixed $r$. Here the mean-field correction to the spectrum takes the following form:
\begin{eqnarray}
\lambda_{k;-}^{(s)}&= &\lambda_{k;-}^{(s=0)} +s\lambda_{k;\epsilon}+O(s^2), \quad k\in\mathbb{Z}
\\
\lambda_{k;\epsilon} &\doteq& \frac{8 r^2 (r-2 \lambda_{k;-}^{(s=0)}  )}{(r-\lambda_{k;-}^{(s=0)}  )^2 (r \tau +4) ((r-2 \lambda_{k;-}^{(s=0)}  ) \tau +4)}.
\end{eqnarray}
One observes that at sufficiently small $r$ and $s>0$, $\Re(\lambda_{k;\epsilon})$ is negative; however, it grows with $r$ to become positive at some critical value, which then leads to acceleration of the relaxation. 

The remaining analysis of the spectrum from Eq.~(\ref{eq:spectrum-}) is computational, and it is focused on testing the effect of the MF nonlinearity. We choose to experiment with $r$ and other parameters in the algebraic functional form, $g(N)=r (2 N)^{s/2}$ while keeping the main characteristic of the ensemble, $\tau$, fixed. Our choice of the functional form is such that the stationary point achieved at different $s$ gives the same $g(1/2)=r$. The dependency over $r$ and $s$ of eigenvalues is shown in Figs.~(\ref{fig:eigen_r}).

%

\section{Direct Numerical Simulations of the Ensemble.}
\begin{figure}[!htp]
            \includegraphics[width=0.48\textwidth]{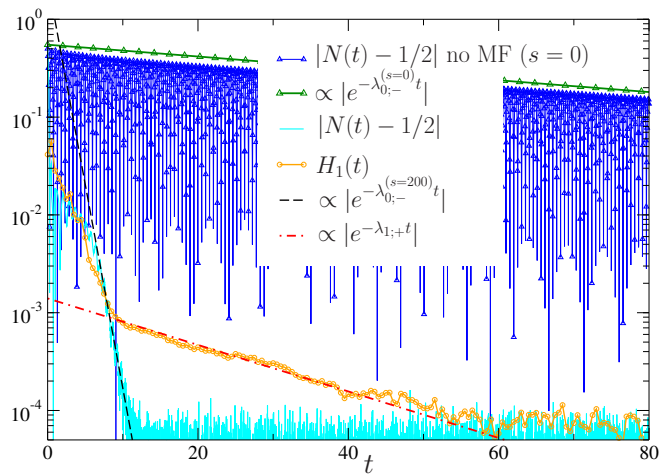}
             \caption{We compare relaxation of $|N_\uparrow(t)-1/2|$ and $H_1(t)$ for $s=200$ in the simulations with $r=100$, $\tau=1$, $-\xd=\xu=1$. $N=2\cdot 10^6$, with the initial conditions described in the text. 
             Consistently with the theoretical analysis,
             $N_\uparrow(t)$ decays to its equilibrium value with a real exponential rate, $\lambda_{0;-}^{(s=200)}=0.93$, while $H_1(t)$ decay is much slower with  $\lambda_{1;+}=0.055- 12 i$ (i.e. super-relaxation). The coefficient $A$ is adjusted to make finite size fluctuations of the ensemble (observed at large $t$) comparable for $|N_\uparrow(t)-1/2|$ and $H_1(t)$. For the same settings and  $s=0$, $|N_\uparrow(t)-1/2|$ decays with the rate $\lambda_{0:-}^{(s=0)}=0.014 - 6.0 i$. Note that in this case, $\lambda_{1:+}$, does not depend on $s$ and results in a sub-leading contribution to the long time asymptotic.} 
    \label{fig:H_decay}
\end{figure}

We also complement analysis of the MFC model with direct numerical simulations.  
We initiate dynamics with the same initial condition (IC) for $P(x|0)$. This setting allows us a fair quantitative comparison of the effect (same IC and final state), measured by $s$, of the mean-field nonlinearity strength on the transient to the steady state in Fig.~\ref{fig:H_decay}.  We experiment with the ensemble of $N=2\cdot 10^6$ identical devices following Eqs.~(\ref{eq:basic_x_model}, \ref{eq:sigma_model}).
In Fig.~(\ref{fig:H_decay}) the initial condition was chosen such that all devices are ``on'', i.e. $\sigma=\uparrow$, and with $x$ distributed uniformly over $[\xd,\xu]$. Results of the experiments are illustrated in Figs.~(\ref{fig:H_decay}). 
These experiments, combined with the analytic analysis, have allowed us to uncover interesting features of the model described in the following.
\begin{figure}[!htp]
            \includegraphics[width=0.48\textwidth]{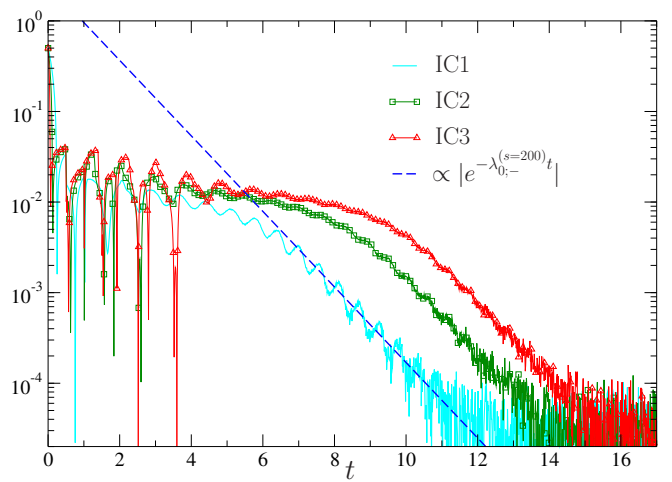}
             \caption{We compare relaxation of $|N_\uparrow(t)-1/2|$ to the stationary state for different strongly non-perturbative Initial Conditions -- IC1, IC2, IC3 described in the text. The simulation parameters are the same as in Fig.~\ref{fig:H_decay}. After a short transient, all cases tested relax to the stationary state with the rate $\lambda_{0;-}^{(s=200)}\approx 0.93$ predicted by linear theory. (Notice that the time span shown in the Figure is the different from Fig.~\ref{fig:H_decay}.)} 
    \label{fig:IC}
\end{figure}

\underline{Significant enhancement of the relaxation by MFC} is 
reported at sufficiently large nonlinearity. For example, comparing $s=200$ and $s=0$ at $r=100$ one observes in Fig.~\ref{fig:H_decay}, $\lambda_{0:-}^{(s=200)}/\Re(\lambda_{0:-}^{(s=0)})\simeq 70$-fold gain in the exponential rate of decay! Moreover, the gain is also accompanied with suppression of oscillations ($\lambda_{0;-}^{(s=200)}$ is real)  and only $N_{\rm{out}}^{\rm{(st)}}\simeq  3.8\%$ of devices are outside their comfort zone.


\underline{Super-Relaxation.} The last,  but not the least, noticeable feature of the model seen in Fig.~(\ref{fig:H_decay}) is that when the MFC is sufficiently large the aggregate consumption of the ensemble relaxes to its steady value much faster than the entire distribution. We choose to compare relaxation of $N_\uparrow(t)$,  corresponding to the total energy consumed by the ensemble with, 
\begin{equation}
    H_1(t)=A \int \sum_{j=\uparrow,\downarrow}|P_{j}(x|t)-P_{j}^{\rm{(st)}}(x)| \,\d x,
\end{equation}
which accounts for all of the ``$\pm$''-modes, contrary to $N_\uparrow(t)$, and is dependent only on the ``$-$''-branch of the spectrum Eq.~\eqref{eq:spectrum-}. 
Figs.~(\ref{fig:H_decay},\ref{fig:eigen_r}) show that the total energy consumption of the ensemble, $N_\uparrow(t)$, relaxes to a constant much faster than $H_1(t)$. This observation is fully consistent with the results of our theoretical analysis, stating that at sufficiently large $s$, $\Re(\lambda^{(s)}_{0;-})$ which controls $N_\uparrow(t)$ relaxation, is larger than $\Re(\lambda_{1;+})$, controlling relaxation of any other expectations of a general position over $P(x|t)$, including $H_1(t)$. 

\underline{Dependence on the initial condition.}
To verify validity of the linear analysis in the regime where initial condition is far from the stationary state, we have conducted independent numerical investigation. We test strongly non-perturbative initial conditions of the following type:  
\begin{itemize}
    \item[(IC1)] 
    All devices are uniformly distributed over $x\in[-1,1]$ and all are on. 
    \item[(IC2)] All devices are uniformly distributed over $x\in[-1,0]$ and all are on. 
    \item[(IC3)] All devices are uniformly distributed over $x\in[-1,-0.5]$ and all on. 
\end{itemize}
The results of the numerical investigation are shown in Fig.~(\ref{fig:IC}). We observe that all ICs converge to the steady state after a short transient period, and the rate of convergence to the steady state is numerically close to predictions of the linear theory. 

Note that we simulate the Fokker-Planck system of Eq.~(\ref{eq:FP},\ref{eq:FP_full}) as finite ensemble of large but finite number of stochastic ``elements''. Fluctuations, associated with the finite size effect are of the standard ``law of large numbers'' type.  For example,
and as clearly seen in Fig.~(\ref{fig:H_decay}), the long time asymptotic  of $|N_\uparrow - 1/2|$ and $H_1$ saturate to a plateau (instead of continuing to decay to zero) which scales as $N^{-1/2}$.
\section{Conclusion}
The main message of this manuscript is in an unexpectedly strong positive effect of the \emph{Mean-Field Nonlinear control} on mixing in the ensemble of energy loads. Our analytic and computational analyses show that making the transition rates between switch-on and switch-off states to depend on the instantaneous number of devices in the two states accelerates mixing in the ensemble significantly, in comparison with the bare case where the rates are constant. The effect is especially well pronounced at a sufficiently strong level of the mean-field nonlinearity when the relaxation of the probability density distribution vector is almost of the pure decay (no oscillations) type. In this regime we also observe and explain the \emph{super-relaxation} effect:  the total number of the switch-on devices (which is also instantaneous energy consumption of the system) decays to its steady state much faster than any other mode of the probability density distribution vector. We also observe and characterize transitions between pure decay and decay + oscillations regimes.

We expect that the majority of features of the ensemble just highlighted are universal and robust to modification of the basic model of the device dynamic. In particular, we link emergence of the Super-Relaxation to the fact that the ``+''-modes do not contribute to the total energy ($\int \rho_{+,\uparrow}(x) \d x=0$) thus not affected by the MFC.

In terms of the future work, we plan to investigate the joint effect of the two complementary strategies of the ensemble mixing enhancement: the MFC studied in this manuscript and control by randomization discussed in our preceding manuscript \cite{18CCD}. Analyzing the effectiveness of the mean-field approach for a broader control objective 
constitutes another research line to explore. In particular, it will be of interest to consider mean-field generalization of the Markov Decision Process model of the energy load ensemble discussed in \cite{18CCD}. 

We conclude expressing our belief that the phenomenon of mixing enhancement by mean field, motivated and discussed in this manuscript in the context of the energy system ensemble, may also apply to a much broader class of natural and engineered systems, such as particles following a turbulent flow, birds in a flock formation, or robots exploring a landscape. In particular, many models of animal behaviour \cite{Eftimie6974,Eftimie2007} and statistical mechanics models describing interacting particles subject to a non-equilibrium flux, such as run-and-tumble particles on the ring \cite{mallmin2018exact}, share mathematical structure with the energy system model analyzed in the manuscript. 

\begin{acknowledgments}
The authors acknowledge useful discussions with Marc Vuffray. The work at LANL was carried out under the auspices of the National Nuclear Security Administration of the U.S. Department of Energy under Contract No. DE-AC52-06NA25396. The work was partially supported by DOE/OE/GMLC and LANL/LDRD/CNLS projects. 
\end{acknowledgments}

\end{document}